\documentclass[
aps,
prl,
reprint,
floatfix,
amsmath,
twocolumn,
superscriptaddress,
]{revtex4-1}

\pdfoutput=1
\usepackage{dcolumn}   
\usepackage{bm}        
\usepackage{amssymb}   
\usepackage{amsfonts}
\usepackage{verbatim}
\usepackage[pdftex]{graphicx}  
\usepackage{amsmath}
\usepackage{amsfonts}
\usepackage{ulem}
\usepackage[pdftex]{color}

\newcommand{\ket}[1]{|#1\rangle}

\def\lsim{\mathrel{\rlap{\lower4pt\hbox{\hskip1pt$\sim$}}
    \raise1pt\hbox{$<$}}}                
\def\gsim{\mathrel{\rlap{\lower4pt\hbox{\hskip1pt$\sim$}}
    \raise1pt\hbox{$>$}}}                

\newcommand\startsupplement{%
    \makeatletter 
       \setcounter{table}{0}
       \renewcommand{\thetable}{S\arabic{table}}
       \setcounter{figure}{0}
       \renewcommand{\thefigure}{S\arabic{figure}}
       \setcounter{equation}{0}
       \renewcommand{\theequation}{S\arabic{equation}}
    \makeatother}

\begin{document}
\normalem

\title{Measurement of a Superconducting Qubit with a Microwave Photon Counter}

\author{A. Opremcak}
\affiliation{Department of Physics, University of Wisconsin-Madison, Madison, Wisconsin 53706, USA}
\author{I. V. Pechenezhskiy}
\altaffiliation[Present address: ]{University of Maryland, College Park, Maryland 20742, USA}
\affiliation{Department of Physics, University of Wisconsin-Madison, Madison, Wisconsin 53706, USA}
\author{C. Howington}
\affiliation{Department of Physics, Syracuse University, Syracuse, New York 13244, USA}
\author{B. G. Christensen}
\affiliation{Department of Physics, University of Wisconsin-Madison, Madison, Wisconsin 53706, USA}
\author{M. A. Beck}
\affiliation{Department of Physics, University of Wisconsin-Madison, Madison, Wisconsin 53706, USA}
\author{E. Leonard Jr.}
\affiliation{Department of Physics, University of Wisconsin-Madison, Madison, Wisconsin 53706, USA}
\author{J. Suttle}
\affiliation{Department of Physics, University of Wisconsin-Madison, Madison, Wisconsin 53706, USA}
\author{C. Wilen}
\affiliation{Department of Physics, University of Wisconsin-Madison, Madison, Wisconsin 53706, USA}
\author{K. Nesterov}
\affiliation{Department of Physics, University of Wisconsin-Madison, Madison, Wisconsin 53706, USA}
\author{G. J. Ribeill}
\altaffiliation[Present address: ]{Raytheon BBN Technologies, Cambridge, Massachusetts 02138, USA}
\author{T. Thorbeck}
\altaffiliation[Present address: ]{IBM T. J. Watson Research Center, Yorktown Heights, New York 10598, USA}
\author{F. Schlenker}
\affiliation{Department of Physics, University of Wisconsin-Madison, Madison, Wisconsin 53706, USA}
\author{M. G. Vavilov}
\affiliation{Department of Physics, University of Wisconsin-Madison, Madison, Wisconsin 53706, USA}
\author{B. L. T. Plourde}
\affiliation{Department of Physics, Syracuse University, Syracuse, New York 13244, USA}
\author{R. McDermott}
\email[Electronic address: ]{rfmcdermott@wisc.edu}
\affiliation{Department of Physics, University of Wisconsin-Madison, Madison, Wisconsin 53706, USA}

\date{\today}

\begin{abstract}
Fast, high-fidelity measurement is a key ingredient for quantum error correction. Conventional approaches to the measurement of superconducting qubits, involving linear amplification of a microwave probe tone followed by heterodyne detection at room temperature, do not scale well to large system sizes. Here we introduce an alternative approach to measurement based on a microwave photon counter. We demonstrate raw single-shot measurement fidelity of 92\%. Moreover, we exploit the intrinsic damping of the counter to extract the energy released by the measurement process, allowing repeated high-fidelity quantum non-demolition measurements. Crucially, our scheme provides access to the classical outcome of projective quantum measurement at the millikelvin stage. In a future system, counter-based measurement could form the basis for a scalable quantum-to-classical interface.
\end{abstract}

\maketitle

In order to harness the tremendous potential of quantum computers, it is necessary to implement robust error correction to combat decoherence of the fragile quantum states. Error correction relies on high-fidelity, repeated measurements of a significant fraction of the quantum array throughout the run time of the algorithm \cite{Fowler12}. In the context of superconducting qubits, measurement is performed by heterodyne detection of a weak microwave probe tone transmitted across or reflected from a linear cavity that is dispersively coupled to the qubit \cite{Blais04, Siddiqi04, Mallet09, Jeffrey14, Kelly15, Martinis15, Sank16}. This approach relies on liberal use of bulky, magnetic nonreciprocal circuit components to isolate the qubit from noisy amplification stages \cite{Johnson12, Jeffrey14, Macklin15}; moreover, the measurement result is only accessible following room temperature heterodyne detection and thresholding, complicating efforts to implement low-latency feedback conditioned on the measurement result \cite{Riste12, Riste13}. The physical footprint, wiring heat load, and latency associated with conventional amplifier-based qubit measurement stand as major impediments to scaling superconducting qubit technology.

\begin{figure}[h!]
    \includegraphics[width=.48\textwidth]{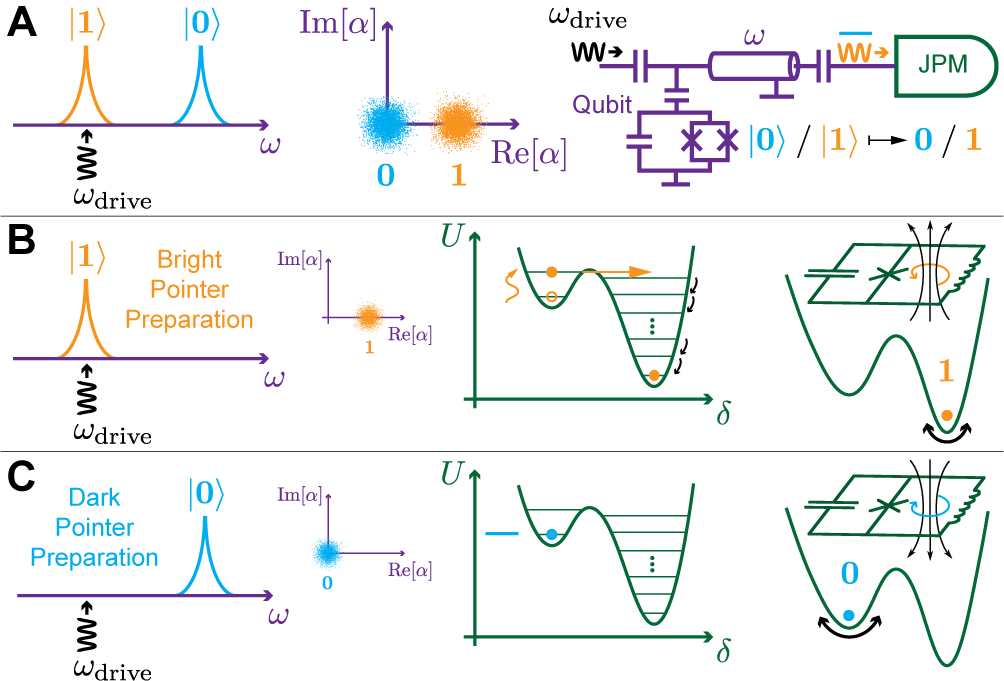}
     \caption{\textbf{Qubit state measurement using the JPM.} (\textbf{A}) Measurement overview. Microwave drive at the dressed cavity resonance corresponding to the qubit $|1\rangle$ state creates bright and dark cavity pointer states with large differential photon occupation. These pointer states are detected using the JPM, which stores the measurement result as a classical bit at the millikelvin stage. (\textbf{B}) Bright pointer detection. Microwaves resonant with the JPM promote the circuit from the ground state of a metastable local minimum (here, left potential well) to an excited state. The detector subsequently undergoes a rapid tunneling transition that allows relaxation to the global minimum of the potential (here, right potential well). (\textbf{C}) Dark pointer detection. Energy contained in the dark pointer state is insufficient to induce a tunneling event. The presence (\textbf{B}) or absence (\textbf{C}) of an interwell tunnelling transition results in classically distinguishable oscillation (circulating current) states in the detector.  \label{fig:fig1}}
\end{figure}

\begin{figure}[ht!]
    \includegraphics[width=.48\textwidth]{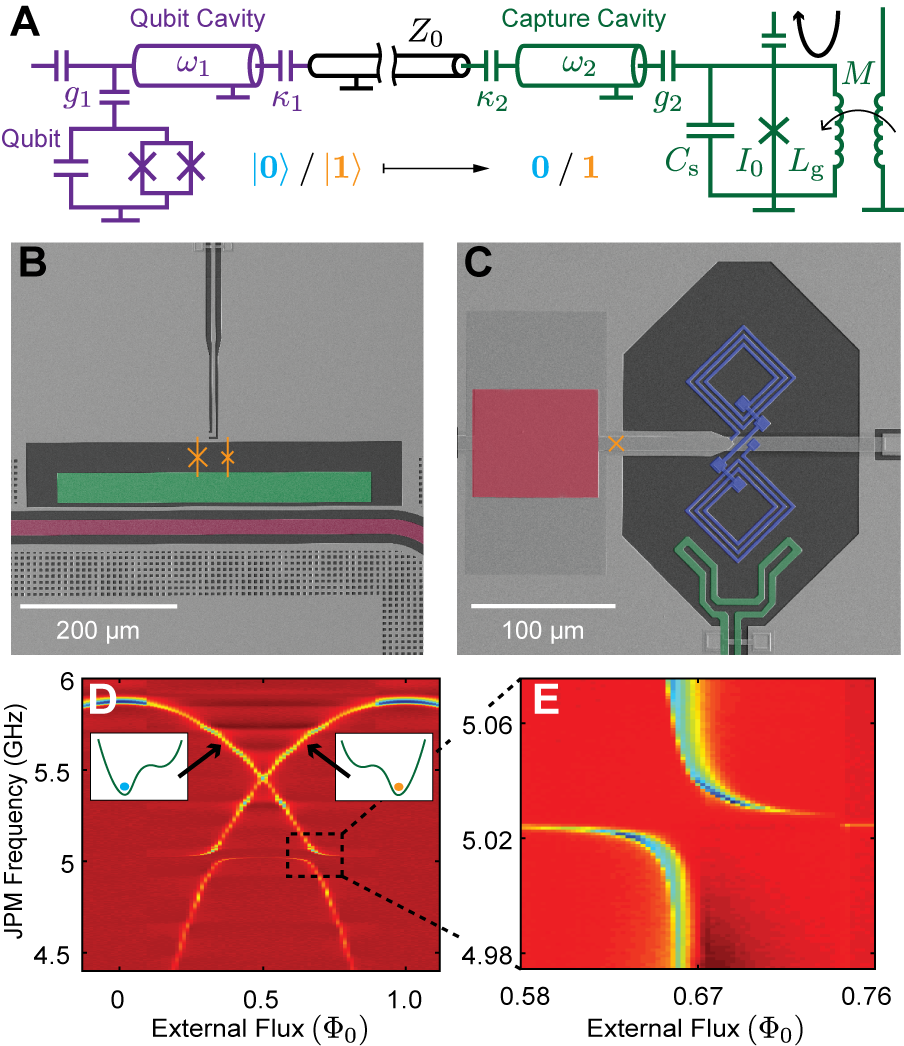}
     \caption{\textbf{Experimental setup.} (\textbf{A}) Circuit schematic. The qubit circuit (purple) is connected to the JPM circuit (green) via a coaxial transmission line (black). (\textbf{B}) Micrograph of the transmon circuit with superconducting island (green), qubit cavity (red), and Josephson junctions (orange). (\textbf{C}) Micrograph of the JPM circuit (capture cavity not shown) with its 3+3 turn gradiometric loop inductance $L_{\text{g}}$ (blue), single Josephson junction with critical current $I_0$ (orange), parallel-plate capacitor $C_{\text{s}}$ (red), and on-chip flux bias line with mutual inductance $M$ (green). (\textbf{D}) JPM spectroscopy versus external flux. Insets show cartoons of a phase particle bound to the left and right wells. (\textbf{E}) Zoom in of the avoided level crossing between the JPM and capture cavity. \label{fig:fig2}}
\end{figure}

An alternative approach involves entanglement of the qubit with the linear resonator to create cavity pointer states characterized by large differential photon occupation, followed by subsequent photodetection \cite{Govia14b}. In our experiments (Fig. \ref{fig:fig1}A), microwave drive at one of the two dressed cavity frequencies maps the qubit state onto ``bright'' and ``dark'' cavity pointer states. Discrimination of the states is performed directly at the millikelvin stage by the Josephson photomultiplier (JPM), a microwave photon counter; we use no nonreciprocal components between the qubit and JPM. The JPM is based on a single Josephson junction in an rf superconducting quantum interference device (SQUID) loop that is biased close to the critical flux where a phase slip occurs. The circuit parameters are chosen to yield a potential energy landscape with one or two local minima, depending on flux bias (see S1 for a theoretical treatment of the JPM); the distinct local minima correspond to classically distinguishable circulating current states in the device. Once the JPM is properly biased, the presence of resonant microwaves induces a rapid tunneling event between the two classically distinguishable states of the detector (Fig. 1B). In the absence of microwave input, transitions occur at an exponentially suppressed dark rate (Fig. 1C). Thus, the absorption of resonant microwaves creates an easily measured ``click" \cite{Chen11}.

A schematic of the experiment is shown in Fig. \ref{fig:fig2}A. The qubit and the JPM are fabricated on different silicon substrates and housed in separate aluminum enclosures connected via a coaxial transmission line with characteristic impedance $Z_0=50~\Omega$ (see S2-S3 for fabrication details). The qubit chip (purple circuit in Fig. \ref{fig:fig2}A) incorporates an asymmetric transmon \cite{JKoch07, Barends13, Hutchings17} that is capacitively coupled to a half-wave coplanar waveguide (CPW) resonator, the qubit cavity, with frequency $\omega_{1}/2\pi = 5.020$ GHz and qubit-cavity coupling strength $g_1/2\pi = 110$ MHz. The qubit anharmonicity $\alpha/2\pi = -250$ MHz. A micrograph of the transmon is shown in Fig. \ref{fig:fig2}B. In our experiments, the qubit is operated at a fixed frequency $\omega_{\text{q}}/2\pi = 4.433$ GHz.

The JPM (green circuit in Fig. \ref{fig:fig2}A) is based on the capacitively-shunted flux-biased phase qubit \cite{Steffen06}. The JPM is capacitively coupled to a local auxiliary CPW resonator, the capture cavity, with bare frequency $\omega_{2}/2\pi = 5.028$ GHz and coupling strength $g_2/2\pi = 40$ MHz. A micrograph of the JPM is shown in Fig. \ref{fig:fig2}C. The circuit involves a single Al-AlO$_x$-Al Josephson junction with critical current $I_0=$ 1~$\mu$A embedded in a 3+3 turn gradiometric loop with inductance $L_\text{g}=$~1.1~nH. The junction is shunted by an external parallel-plate capacitor $C_\text{s}=$~2~pF. The plasma frequency of the JPM is tunable with external flux from $5.9$ GHz to $4.4$ GHz (Fig. \ref{fig:fig2}D-E), allowing for both resonant and dispersive interactions between the JPM and capture cavity.

The qubit and capture cavities are coupled to the mediating transmission line with leakage rates $\kappa_1= 1/(260 \> \text{ns})~\text{and}~\kappa_2 = 1/(40 \> \text{ns})$, respectively; there are no intervening isolators or circulators (see S4 for details on the experimental setup). Following pointer state preparation, microwave energy leaks out of the qubit cavity and a fraction of that energy is transferred to the capture cavity. In the absence of dynamic tuning of the capture cavity, the maximum transfer efficiency is $4/e^2\approx$~54\% for cavities that are perfectly-matched in frequency and decay rate \cite{Wenner14}. Given the mismatch in decay rates of our qubit and capture cavities, we expect transfer efficiency around 30\% for cavities that are well matched in frequency (see S5 for an analysis of pointer state transfer efficiency).

\begin{figure*}[t]
    \includegraphics[scale = 1.0]{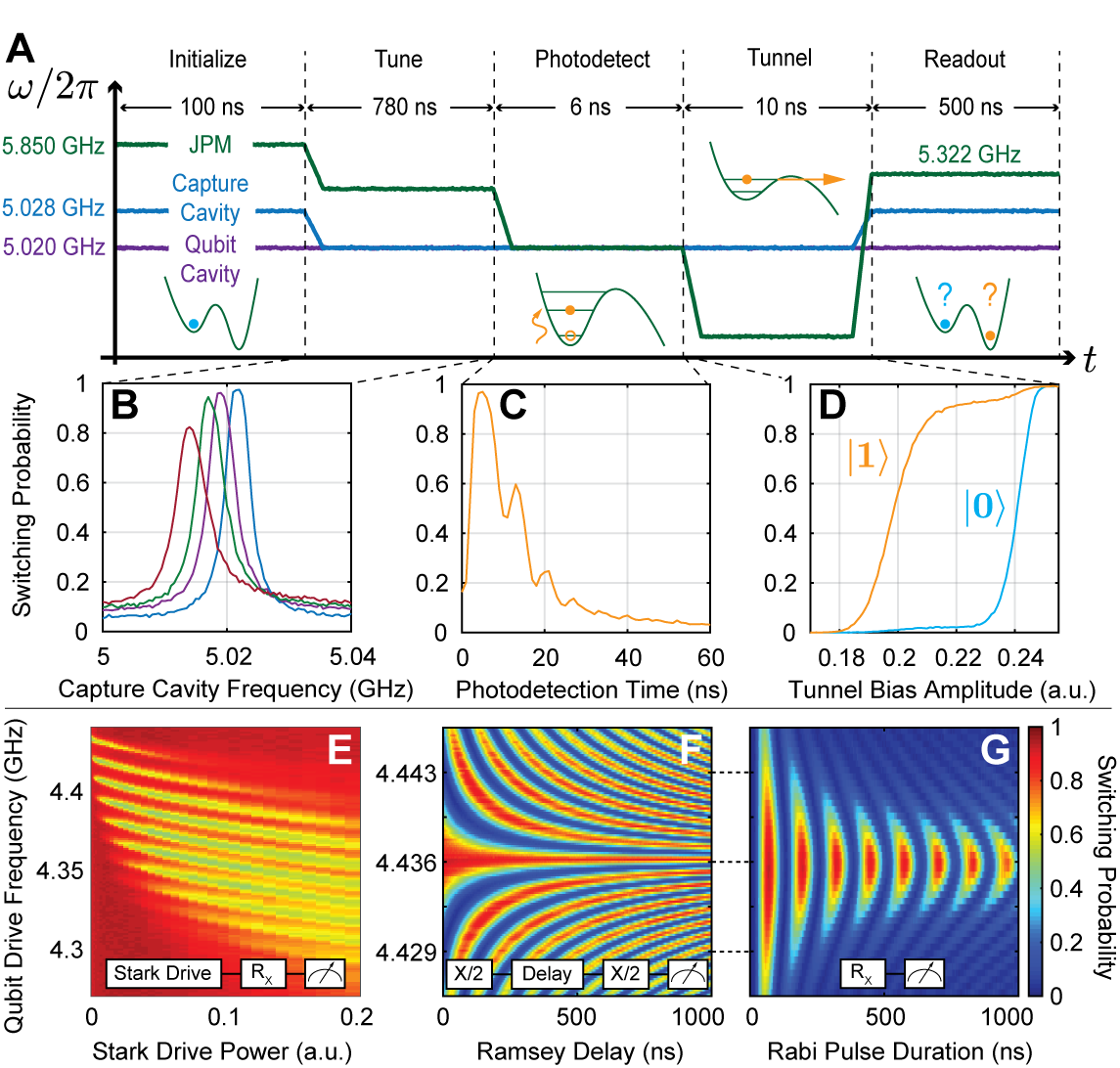}
     \caption{\textbf{JPM-based qubit measurement.} (\textbf{A}) Timing diagram. The JPM is first initialized in a well-defined flux state. (\textbf{B}) The bias of the JPM is then adjusted to fine tune the effective frequency of the capture cavity. During the ``tune" interval, bright or dark pointer states leak from the qubit cavity to the capture cavity. (\textbf{C}) Next, the JPM is tuned to resonance with the capture cavity to probe for any captured photons. When present, photons are transferred from the capture cavity to the JPM, inducing intrawell transitions. (\textbf{D}) Finally, a brief bias pulse is applied to induce tunneling of excited JPM states. At this point the measurement is complete. Microwave reflectometry is subsequently used to interrogate the JPM. (\textbf{E}) Stark spectroscopy used to calibrate qubit cavity photon occupation. (\textbf{F}) JPM-detected Ramsey fringes versus qubit drive detuning. (\textbf{G}) JPM-detected Rabi oscillations versus qubit drive detuning. \label{fig:fig3}}
\end{figure*}

A timing diagram of the measurement is shown in Fig. \ref{fig:fig3}A; the cartoon insets depict the dynamics of the JPM phase particle at critical points throughout the measurement sequence. We begin by initializing the JPM in one of the two local minima of its potential. The bias point of the JPM is then adjusted to tune the capture cavity resonance in order to maximize photon transfer efficiency (Fig. \ref{fig:fig3}B). Following qubit operations, a drive tone is applied to the qubit cavity that maximizes intensity contrast between microwave cavity pointer states (not shown; see S6 for details on drive tone pulse shaping). In our experiments, the bright pointer state corresponds to a mean qubit cavity photon occupation $\bar{n}\sim10$, calibrated using the ac Stark effect (Fig. \ref{fig:fig3}E; see S7 for qubit and capture cavity photon occupation estimates) \cite{Schuster05, Schuster07}. The microwave pointer states leak from the qubit cavity to the capture cavity on a timescale $\sim1/\kappa_1$. After pointer state transfer, the JPM is biased into resonance with the capture cavity and occupation of that mode induces intrawell excitations of the phase particle on a timescale $\pi/2 g_2\sim$~6~ns (Fig. \ref{fig:fig3}C) \cite{Hofheinz08}. Finally, a short ($\sim$10~ns) bias pulse is applied to the JPM to induce interwell tunneling of excited states \cite{Cooper04}; the amplitude of the bias pulse is adjusted to maximize tunneling contrast between qubit excited and ground states (see Fig. \ref{fig:fig3}D). At this point the measurement is complete: the measurement result is stored in the classical circulating current state of the JPM. To retrieve the result of qubit measurement for subsequent analysis at room temperature, we use a weak microwave probe tone to interrogate the plasma resonance of the JPM following measurement. JPM bias is adjusted so that the plasma frequencies associated with the two local minima in the potential are slightly different; reflection from the JPM can distinguish the circulating current state of the detector with $>$ 99.9\% fidelity in under $500$~ns (see S4 for details on JPM state interrogation).

Each measurement cycle yields a binary result -- ``0" or ``1" -- the classical result of projective quantum measurement. To access qubit state occupation probabilities, the measurement is repeated many times. The JPM switching probabilities represent raw measurement outcomes, uncorrected for state preparation, relaxation, or gate errors. In Fig. \ref{fig:fig3}F and \ref{fig:fig3}G we display the raw measurement outcomes for qubit Ramsey and Rabi experiments, respectively. The JPM measurements reported here achieve a raw fidelity of $92\%$. The bulk of our fidelity loss is due to qubit energy relaxation during pointer state preparation and dark counts, which contribute infidelity of 5\% and 2\%, respectively. In our setup, dark counts stem from both excess $|1\rangle$ population of the qubit and spurious microwave energy contained in our dark pointer state. We attribute the remaining infidelity to imperfect gating and photon loss during pointer state transfer. Improved cavity matching would enable JPM operation at deeper bias points, leading to a suppression of dark counts. Similarly, for a fixed qubit energy relaxation time $T_1$, relaxation errors could be suppressed by moving to leakier cavities, speeding the transfer of photons from the qubit cavity to the capture cavity. 

\begin{figure*}[t]
    \includegraphics[scale = 1.0]{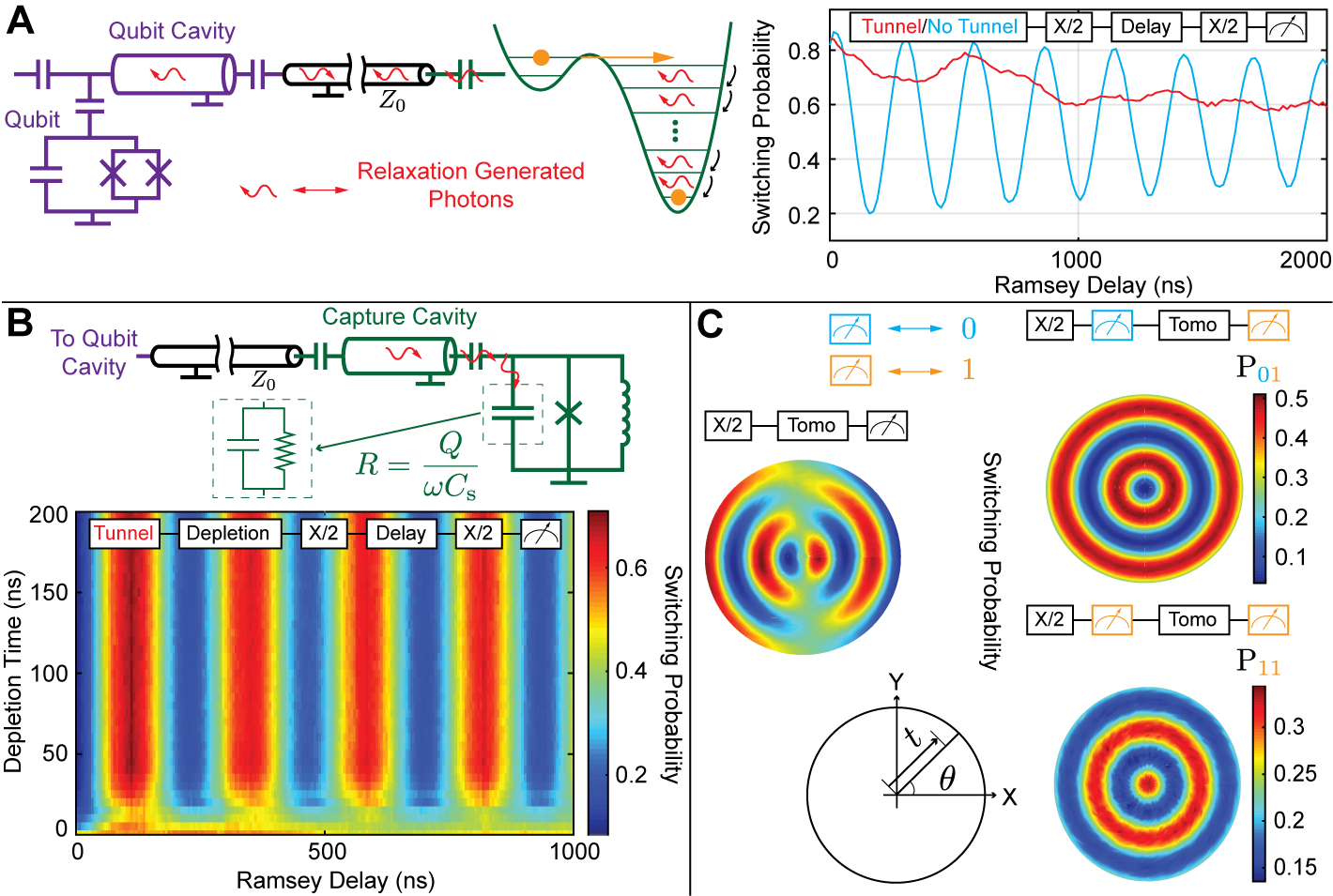}
     \caption{\textbf{Mitigating backaction and preserving QNDness.} (\textbf{A}) The JPM switching event releases energy of order 100 photons, inducing spurious population of the capture and qubit cavities. The right-hand panel shows baseline qubit Ramsey fringes (blue) and Ramsey fringes measured after a forced tunneling event by the JPM (red). (\textbf{B}) Following the fast flux pulse that induces JPM tunneling, we adjust JPM bias so that energy deposited in the cavities is dissipated in the JPM, yielding a deterministic reset of the cavities. The color plot shows qubit Ramsey fringes versus duration of the depletion interaction between the JPM and the capture cavity. (\textbf{C}) Qubit tomography following JPM-based measurement. We prepare the superposition state $(\ket{0}-i\ket{1})/\sqrt{2}$ and verify the state with overdetermined qubit tomography (left panel). To characterize the qubit state after JPM-based measurement, we prepare the same superposition state, measure with the JPM, and then perform qubit tomography on the resulting state. Qubit tomography conditioned on the JPM measurement shows high overlap with target states $|0\rangle$ (top right) and $|1\rangle$ (bottom right). \label{fig:fig4}}
\end{figure*}

JPM-based photodetection of the bright pointer state involves release of a large energy of order 100 photons as the JPM relaxes from a metastable minimum to the global minimum of its potential \cite{McDermott05}. It is critical to understand the backaction of JPM-based measurement on the qubit state. The JPM tunneling transient has a broad spectral content, and Fourier components of this transient that are resonant with the capture and qubit cavities will induce a spurious population in these modes that will lead to photon shot noise dephasing of the qubit \cite{Sears12, Yan16}. In Fig. \ref{fig:fig4}A we show the results of qubit Ramsey scans performed with (red) and without (blue) a forced JPM tunneling event prior to the experiment. In the absence of any mitigation of the classical backaction, qubit Ramsey fringes show strongly suppressed coherence and a frequency shift indicating spurious photon occupation in the qubit cavity \cite{McClure16}. However, we can use the intrinsic damping of the JPM mode itself to controllably dissipate the energy in the linear cavities and fully suppress photon shot noise dephasing. Immediately following JPM-based measurement, the JPM is biased to a point where the levels in the shallow minimum are resonant with the linear cavity modes. Energy from the capture cavity leaks back to the JPM, inducing intrawell transitions; at the selected bias point, the interwell transition probability is negligible. The JPM mode is strongly damped, with quality factor $Q\sim300$, set by the loss tangent of the SiO$_2$ dielectric used in the JPM shunt capacitor \cite{Martinis05}. As a result, the energy coupled to the JPM is rapidly dissipated. With this deterministic reset of the cavities, fully coherent qubit Ramsey fringes are recovered for depletion times $\gsim$~40~ns as shown in Fig. \ref{fig:fig4}B. We reiterate that no nonreciprocal components are used in these experiments to isolate the qubit chip from the classical backaction of the JPM.

In Fig. \ref{fig:fig4}C we explore the quantum non-demolition (QND) character of our measurement protocol \cite{Lupascu04}. We prepare the qubit in the superposition state $(\ket{0}-i\ket{1})/ \sqrt{2}$ aligned along the $-y$ axis of the Bloch sphere. We verify the state by performing overdetermined tomography after \cite{Steffen06}. Here the direction $\theta$ and length $t$ of a tomographic pulse are swept continuously over the equatorial plane of the Bloch sphere prior to measurement. For control pulses applied along the $x$-axis, the qubit undergoes the usual Rabi oscillations; for control applied along $y$, the qubit state vector is unaffected. Following an initial JPM-based measurement (including depletion interaction and additional delay for qubit cavity ringdown), we perform a tomographic reconstruction of the qubit state by applying a pre-rotation and a final JPM-based measurement. In the right-hand panel of Fig. \ref{fig:fig4}C, we display tomograms corresponding to the classical measurement results ``0" (top) and ``1" (bottom). When the measurement result ``0" is returned, we find a tomogram that overlaps with the ideal $\ket{0}$ state with fidelity 91\% (see S8 for details on extracting overlap fidelity from the tomograms). When the result ``1" is returned, the measured tomogram corresponds to overlap fidelity of 69\% with the $\ket{1}$ state. The loss in fidelity for the qubit $\ket{1}$ state is consistent with the measured qubit $T_1$ time of 6.6~$\mu$s and the 2.8~$\mu$s measurement cycle. We conclude that our JPM-based measurement is highly QND.

In conclusion, we have demonstrated a high-fidelity, fast photon counter-based qubit measurement. The approach provides access to the binary result of projective quantum measurement at the millikelvin stage and could form the basis of the measurement side of a robust, scalable interface between a quantum array and a proximal classical controller based on the single flux quantum digital logic family \cite{Fedorov14, McDermott18}.

\medskip

\begin{acknowledgments}
We acknowledge stimulating discussions with Mike Vinje. Portions of this work were performed in the Wisconsin Center for Applied Microelectronics, a research core facility managed by the College of Engineering and supported by the University of Wisconsin - Madison. Other portions were performed at the Cornell NanoScale Facility, a member of the National Nanotechnology Coordinated Infrastructure (NNCI), which is supported by the National Science Foundation under Grant No. ECCS-1542081. This work was supported by the U.S. Government under ARO Grants W911NF-14-1-0080 and W911NF-15-1-0248.

A.O. and I.V.P. contributed equally to this work.
\end{acknowledgments}

\bibliographystyle{unsrtRFM}
\bibliography{master_refs_Jun17}
\clearpage

\startsupplement
\part{Supplementary Information}
\section{S1. JPM Theory}
\begin{figure}[h!]
    \includegraphics[width=0.45\textwidth]{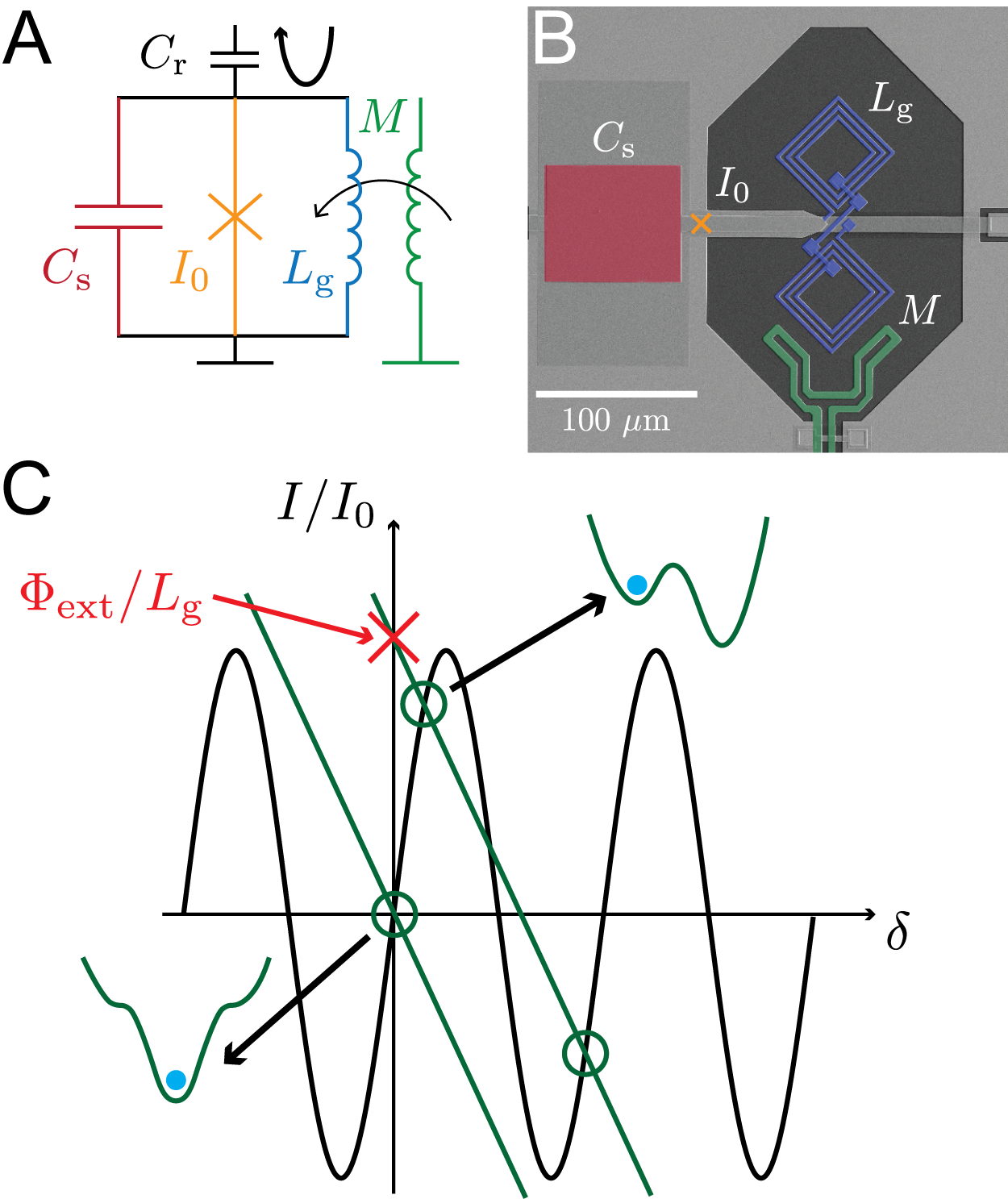}
     \caption{JPM design. \textbf{(A)} Circuit schematic of the JPM. \textbf{(B)} SEM micrograph of the device. Components are color coded to match the schematic (the JPM reflection capacitor $C_{\text{r}}$ is not shown). \textbf{(C)} Graphical solution of Eq. 2. The slope of the line, -1/$\beta_L$, determines the number of local minima (shown as open circles) for a fixed $\Phi_{\text{ext}}$. External flux $\Phi_{\text{ext}}$ controls the $y$-intercept, allowing us to move between a single- and double-well regime as needed for JPM reset and photodetection. Black arrows show JPM potentials (with phase particles in blue) for two values of $\Phi_{\text{ext}}$. \label{fig:figS1}}
\end{figure}

A circuit schematic for the JPM is shown in Fig. \ref{fig:figS1}A. The JPM is based on the design of the capacitively-shunted flux-biased phase qubit \cite{Steffen06}. Fig. \ref{fig:figS1}B shows a scanning electron microscopy (SEM) micrograph of the circuit with labels indicating components. The circuit Hamiltonian is given by
\begin{equation}
H(\delta, Q) = \frac{Q^{2}}{2C_{\text{s}}} - E_{J} \cos\delta +\frac{1}{2L_{\text{g}}}\left(\frac{\Phi_{0}}{2\pi}\right)^2\left(\delta-\frac{2\pi \Phi_{\text{ext}}}{\Phi_{0}}\right)^2,
\end{equation}
where $Q$ is the capacitor charge, $C_{\text{s}}$ is the shunt capacitance (red), $\delta$ is the phase difference across the Josephson junction, $\Phi_0 \equiv h/2e$ is the magnetic flux quantum, $I_{0}$ is the critical current of the Josephson junction (orange), $E_J=I_{0}\Phi_{0}/2\pi$ is the Josephson energy, and $L_{\text{g}}$ is the gradiometric loop inductance (blue). The capacitance of the Josephson junction is negligible compared to $C_{\text{s}}$. The external flux $\Phi_{\text{ext}}$ is generated by an on-chip control line (green) which is coupled to the JPM with a mutual inductance $M$. The extrema of the potential energy landscape are determined by the equation
\begin{equation}
\sin\delta = \frac{1}{\beta_{L}} \left(\frac{2\pi \Phi_{\text{ext}}}{\Phi_{0}}-\delta\right),
\label{eqn:cc}
\end{equation}
where
\begin{equation}
\beta_{L} = \frac{2\pi L_{\text{g}} I_{0}}{\Phi_{0}}.
\end{equation}
Eq. \ref{eqn:cc} is a straightforward statement of current conservation in the JPM loop; solutions can be depicted graphically as shown in Fig. \ref{fig:figS1}C. We seek values $\beta_{L}$ which allow the JPM to be tuned between a single- and double-well regime for reset and photodetection, respectively. The curvature at the local minima of the potential determines the plasma frequency:
\begin{equation}
\omega_p =\frac{2\pi}{\Phi_0} \bigg[\frac{1}{C_{\text{s}}}\frac{\partial^2 U}{\partial \delta^2}\bigg]^{1/2}.
\end{equation}
In addition, we can estimate the number of levels in a well by $n \approx \Delta U/\hbar\omega_p$, where $\Delta U$ is the potential energy barrier height.

\begin{figure*}[h!]
    \includegraphics[width=0.85\textwidth]{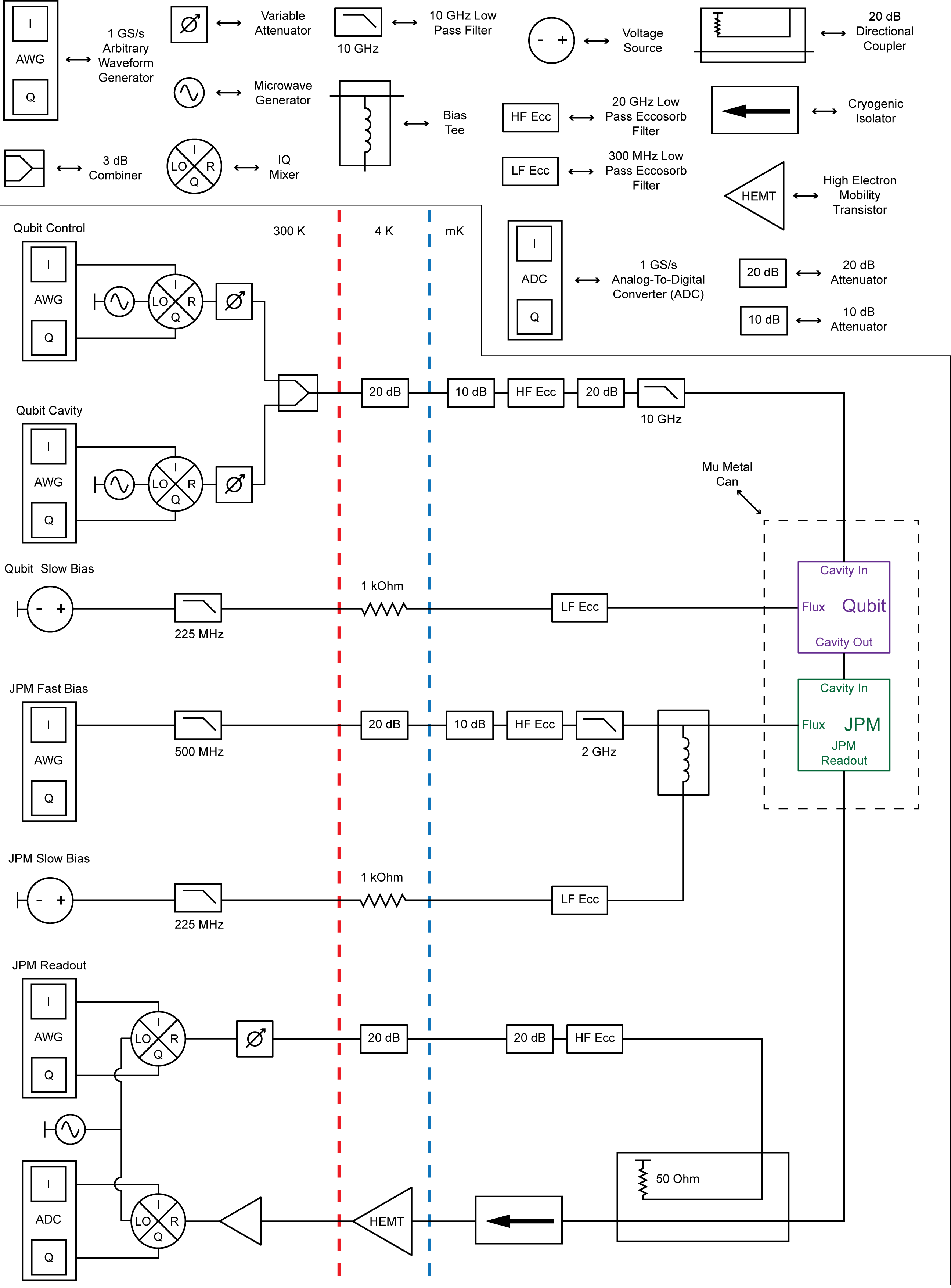}
     \caption{Experimental setup. Dashed colored lines divide temperature stages. Circuit symbols are defined above. Names above the room temperature AWGs and voltage sources describe their role in the experiment. \label{fig:figS2}}
\end{figure*}

\section{S2. JPM Fabrication}
The JPM is fabricated on a high-resitivity Si substrate. Prior to deposition, the wafer is dipped in HF acid to remove native oxide from the surface. Next, we quickly transfer ($\sim1~\text{min}$) the device into a high vacuum (HV) sputter tool to deposit a 100 nm-thick film of Al. The first patterning step defines all Al features except for the wiring crossovers, Josephson junction, and shunt capacitor (see Fig. \ref{fig:figS1}B). The pattern is wet-etched using Transene Aluminum Etchant Type A. After this, a 130 nm-thick film of amorphous SiO$_2$ is deposited using plasma-enhanced chemical vapor deposition (PECVD). Next, we define a $1~\mu\text{m}^2$ via in the dielectric which determines the location of our Josephson junction. Josephson junctions are formed in the sputter tool using the following steps: i) \textit{in situ} ion mill to remove native oxide, ii) controlled oxidation in pure O$_2$ at room temperature (\textit{P}$_{\text{O}_2}$~$\sim$~10 mTorr), and iii) deposition of the Al counterelectrode ($\sim$150 nm thick). The counterelectrode layer is then patterned and etched using the same Al etching procedure as before. Next, we pattern for dielectric removal using a reactive-ion etcher (RIE). Dielectric is cleared throughout except where needed for wiring insulation. A final Al wiring step is completed using liftoff and e-beam evaporation in a separate HV system. Once again an \textit{in situ} ion mill is used to ensure good metal-to-metal contact, a $5~\text{nm}$ layer of Ti is evaporated to promote adhesion, then a 150 nm-thick film of Al is evaporated and the metal is lifted off. This completes the device.

\section{S3. Qubit Fabrication}
The transmon qubit and readout cavity are fabricated on a high-resistivity Si substrate. A 90-nm Nb film is deposited using a dc sputter system. A single photolithography step defines all features, except for the Josephson junctions. This pattern is etched using a reactive-ion etcher. The Dolan-bridge qubit junctions are defined in an MMA/PMMA bilayer exposed on an electron-beam writer. The junctions are deposited in the following steps: i) \textit{in situ} Ar ion mill to remove native oxide from underlying Nb, ii) electron-beam evaporation of 35 nm of Al at +11.5 degrees, iii) controlled oxidation, and iv) 65 nm Al deposition at -11.5 degrees.

\section{S4. Experimental Setup}
\begin{figure}[h!]
    \includegraphics[width=0.45\textwidth]{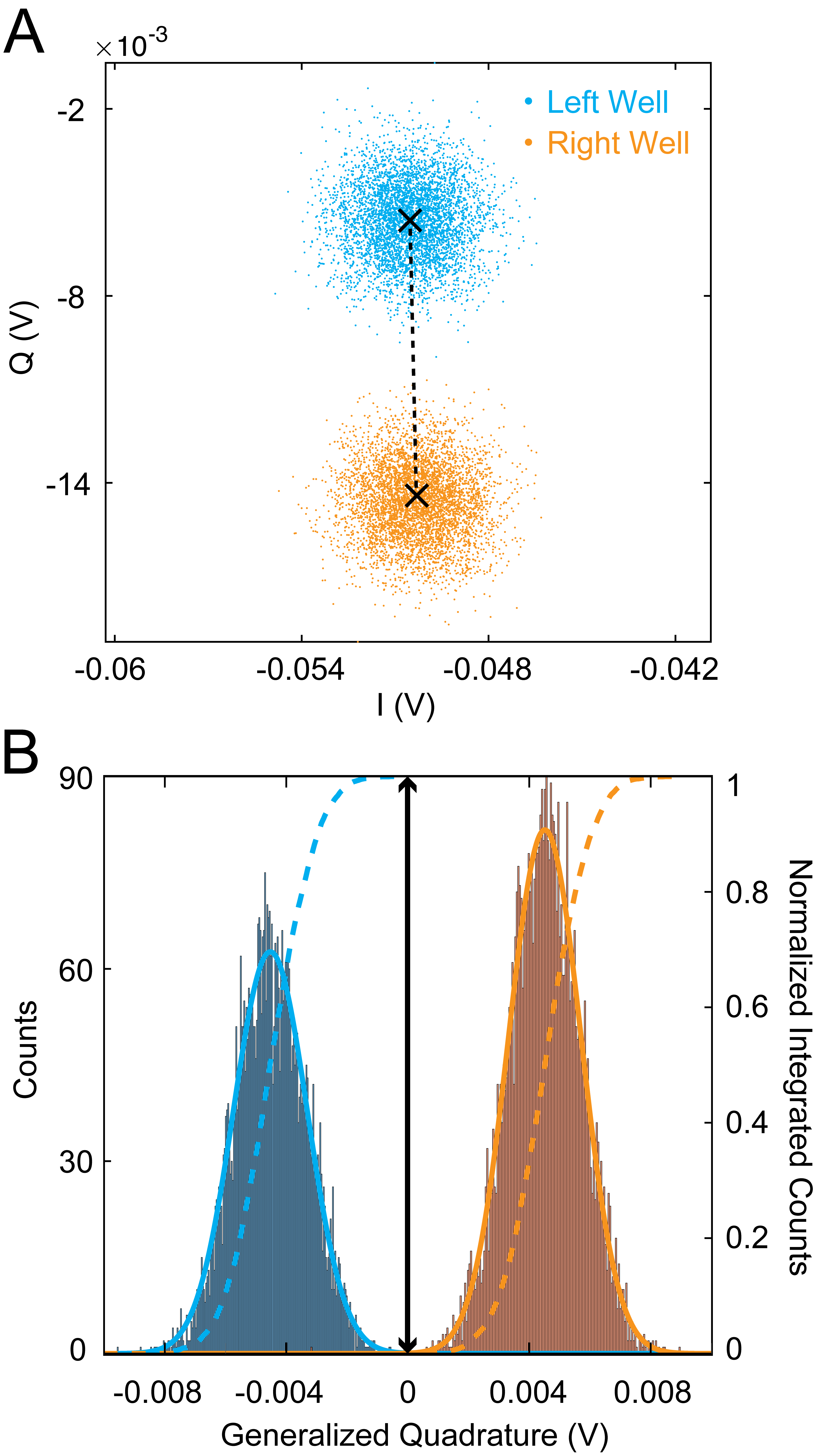}
     \caption{Interrogation of JPM oscillation state. \textbf{(A)} Quadrature amplitudes measured in reflection from a JPM prepared in the two classically distinguishable oscillation states. Single-shot measurement results are projected along the line joining the centroids of the two distributions for the purposes of thresholding. \textbf{(B)} Histograms of the JPM readout results. Solid lines are Gaussian fits, and dashed lines are integrated histograms. Thresholding (double arrow) yields a single-shot fidelity of 99.9\%; the separation fidelity \cite{Jeffrey14} is 99.98\%.\label{fig:figS3}}
\end{figure}

The setup for our experiment is shown in Fig. \ref{fig:figS2}. The qubit control, qubit cavity, and JPM readout waveforms are generated through sideband mixing of shaped intermediate frequency (IF) and local oscillator (LO) tones; 1 GS/s arbitrary waveform generators (AWGs) are used to generate the IF waveforms. These IF waveforms are sent to the in-phase (I) and quadrature (Q) ports of an IQ mixer and are mixed with an LO to generate pulses at microwave frequencies at the RF port. The qubit flux bias is fixed at a constant dc value throughout the measurement sequence. The JPM flux bias is composed of two signals which are combined at the millikelvin stage using a microwave bias tee that is dc coupled to both of its ports. The output of the qubit cavity is connected to the input of the JPM capture cavity via a coaxial transmission line with no intervening isolators or circulators. The state of the JPM is read out in reflection using a directional coupler, an isolator, and a high mobility electron transistor (HEMT) amplifier at the 3 K stage of the cryostat. The JPM readout signal is sent to the RF port of an IQ mixer where it is down-converted using the shared LO with the JPM readout AWG. Baseband I and Q signals are digitized using a 500 MS/s analog-to-digital converter (ADC). Further signal demodulation and thresholding are performed in software in order to extract the oscillation state of the JPM. In Fig. \ref{fig:figS3}A we show our ability to distinguish between distinct oscillation states in IQ space. The JPM state can be determined with $>$ 99.9\% accuracy in under 500 ns (see Fig. \ref{fig:figS3}B).

\section{S5. Pointer State Transfer}
\begin{figure}[ht!]
\includegraphics[width=.45\textwidth]{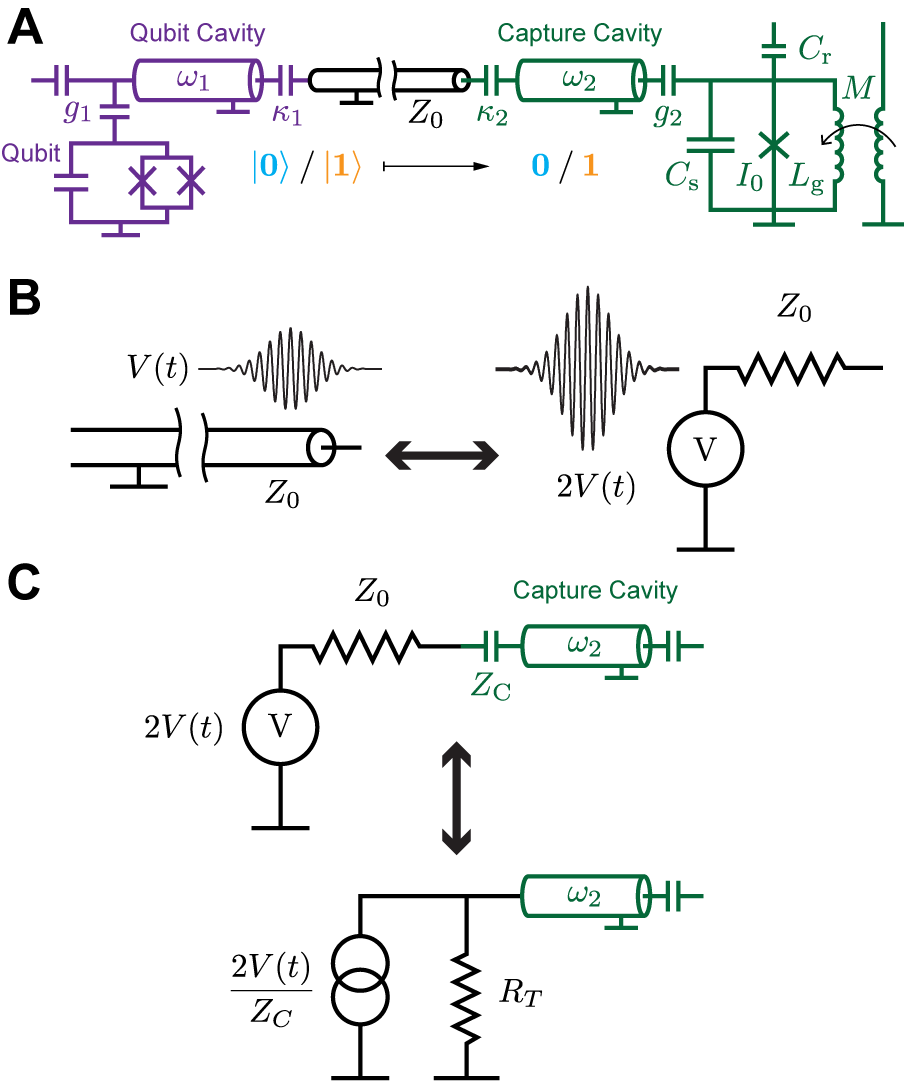}
\caption{Photon transfer from qubit cavity to capture cavity. \textbf{(A)} Schematic for resonator-mediated JPM readout of a qubit. Mode 1 is the dispersively-coupled qubit cavity, fabricated on the same chip as the qubit; mode 2 is the capture cavity fabricated on the same chip as the JPM; and the two chips are joined by a transmission line of arbitrary length. The protocol requires efficient transfer of photons from mode 1 to mode 2, followed by photodetection with the JPM. \textbf{(B)} Thevenin equivalent for a voltage waveform propagating on a transmission line. \textbf{(C)} Thevenin-Norton mapping of the driven photon capture cavity.}
\label{fig:figS4}\end{figure}
Prior work has shown the efficient transfer of photons from resonator to resonator over a transmission line \cite{Wenner14}. These experiments relied on accurate tuning not only of the cavity frequencies and coupling rates, but also of the temporal profile of the cavity coupling rates to the mediating transmission line, in order to achieve transfer efficiencies approaching 100\%. However, as we show in the following analysis, reasonably high ($\sim$ 50\%) transfer efficiency from the qubit cavity to the JPM capture cavity can be achieved with minimal \textit{in situ} tuning, and indeed without any additional tuning elements apart from the JPM itself. 

Our analysis is based on elementary circuit and linear response theory; an alternative picture of efficient photon transfer in terms of destructive interference of reflected waveforms is presented in the Supplement to \cite{Wenner14}.

We consider the experimental setup shown in Fig. \ref{fig:figS4}A. On one chip, the qubit cavity (with frequency $\omega_1$) is coupled to a transmon qubit with strength $g_1$, while on a second chip, the capture cavity (with frequency $\omega_2$) is coupled to a JPM with strenght $g_2$. The two chips are connected by a transmission line of arbitrary length with characteristic impedance $Z_0=50~\Omega$. The qubit cavity is strongly coupled to the output port with a photon number decay rate $\kappa_1$, and the capture cavity is coupled to the transmission line with rate $\kappa_2$. We assume that we can create high-contrast pointer states in the qubit cavity on a timescale much less than $1/\kappa_1$, using the protocol of Govia \textit{et al.} \cite{Govia14b}. The bright pointer state leaks out into the transmission line, leading to a propagating voltage wave:

\begin{align}
V(t) = V_0 e^{-\kappa_1 t/2} \, e^{i\omega_1t}
\end{align}
(for simplicity we omit the trivial spatial dependence of the propagating wave).

At any point on the line, we can exactly model the transmission line as a Thevenin source with voltage $2V(t)$ in series with a Thevenin impedance $Z_0$. In the case where the line is terminated with a matched resistance $Z_0$, all of the energy in the emitted pulse is coupled to the resistor. The maximum available energy is given by
\begin{align}
\frac{V_0^2}{2Z_0}\int_0^\infty e^{-\kappa_1 t} dt = \frac{V_0^2}{2\kappa_1Z_0}.
\label{eqn:en}
\end{align}

Here we calculate the energy that is transferred to the capture cavity as a function of time, and we explore the dependence of photon transfer efficiency on frequency $\omega_2$ and coupling strength $\kappa_2$ of the capture cavity. The drive waveform is replaced by its Thevenin equivalent; the electrical circuit is shown in Fig. \ref{fig:figS4}B. This drive waveform is coupled to the capture cavity via a capacitance with impedance of magnitude $Z_C$, as shown in Fig. \ref{fig:figS4}C. Note that we have $\kappa_2 = Z_0/Z_C^2C$, where $C$ is the effective capacitance of the capture cavity. It is helpful to reexpress the series combination of drive source and coupling capacitor as a Norton equivalent current source, as shown in the lower panel of Fig. \ref{fig:figS4}C. Here, $R_T \equiv Z_C^2/Z_0$ is the transformed impedance of the drive line, and we neglect the slight renormalization of the frequency of the mode.  We use linear response theory to evaluate the energy stored in mode 2 (the capture cavity) following an arbitrary time-dependent drive. The response of mode 2 to a current impulse $I=Q\delta(t)$ is given by
\begin{align}
V_\delta(t) = \frac{Q}{C}e^{-\kappa_2 t/2} \cos\left(\omega_2 t\right).
\end{align}
The mode rings at its resonant frequency, and energy leaks back out to the drive line with rate $\kappa_2$ (we are neglecting internal losses in the cavity). For an arbitrary current drive $I(t)$, the voltage in mode 2 is given by
\begin{align}
V_2(t) = \frac{1}{C} \int_{-\infty}^t I(\tau) e^{-\frac{\kappa_2}{2}(t-\tau)} \cos\left[\omega_2 (t-\tau)\right] d\tau.
\end{align}
For the specific case of our propagating waveform resulting from the decay of mode 1 (the qubit cavity) into the transmission line, we find
\begin{align}
\nonumber
V_2(t) = & \, 2V_0\sqrt{\frac{\kappa_2}{Z_0C}}e^{-\kappa_2t/2} \\
& \, \times \int_0^t e^{\left(\frac{\Delta\kappa}{2}+i\omega_1\right)\tau}\cos\left[\omega_2(t-\tau)\right] d\tau,
\end{align}
where we have introduced the notation $\Delta \kappa \equiv \kappa_2-\kappa_1$.

\begin{figure}[h!]
\includegraphics[width=.45\textwidth]{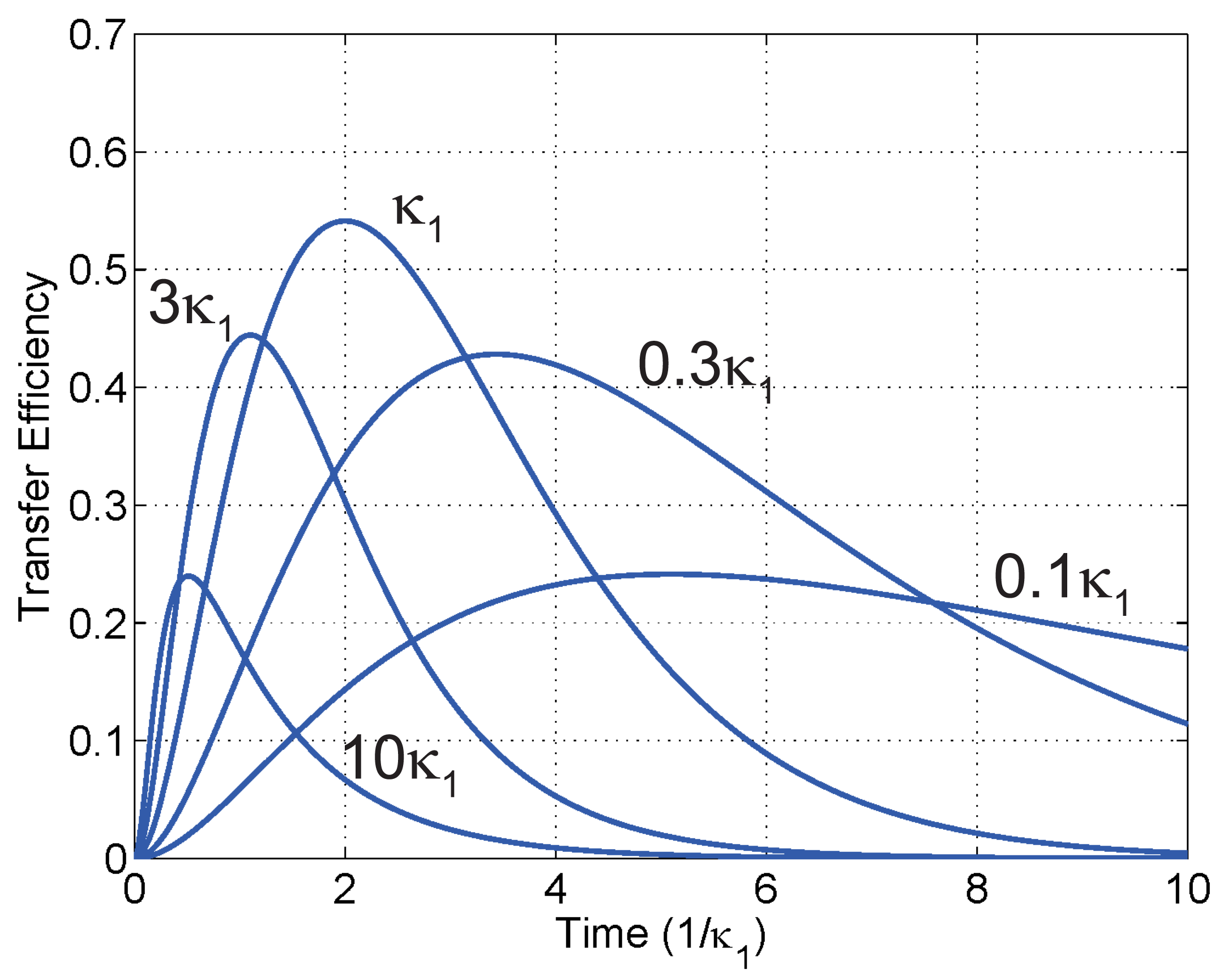}
\vspace*{-0.0in} \caption{Effect of temporal mode mismatch $\kappa_2 \neq \kappa_1$ on photon transfer efficiency. Here the two modes are taken to be on resonance $\omega_1=\omega_2$. The curves are labeled according to the value of $\kappa_2$. The peak transfer efficiency of $4/e^2 \approx$ 54\% is reached for the matching condition $\kappa_2 = \kappa_1$. Even for an order of magnitude mismatch in $\kappa$, transfer efficiencies around 25\% can be obtained.}
\label{fig:kappa}\end{figure}

Let's assume first that the drive frequency and the frequency of mode 2 are matched, $\omega_1=\omega_2\equiv \omega$. In this case, we find
\begin{align}
V_2(t) &= 2V_0\sqrt{\frac{\kappa_2}{Z_0C}}e^{(i\omega - \kappa_2/2)t}\frac{1}{\Delta \kappa}\left(e^{\frac{\Delta\kappa}{2} t} - 1\right).
\end{align}
For vanishing temporal mode mismatch $\kappa_1=\kappa_2\equiv\kappa$, we find
\begin{align}
V_2(t) = V_0\sqrt{\frac{\kappa}{Z_0C}}e^{(i\omega - \frac{\kappa}{2})t}\,t.
\end{align}

The stored energy in mode 2 is given by
\begin{align}
E_2(t) &= \frac{V_0^2}{2Z_0} \kappa e^{-\kappa t} \,t^2.
\end{align}
At time $t_{opt}=2/\kappa$, the stored energy is maximum, with a value
\begin{align}
E_{2,max} = \frac{2}{e^2}\frac{V_0^2}{\kappa Z_0}.
\end{align}
Comparison with Eq. \ref{eqn:en} above shows that the energy has been transferred from mode 1 to mode 2 with efficiency $4/e^2 \approx$ 54\%.

\begin{figure}[h!]
\includegraphics[width=.45\textwidth]{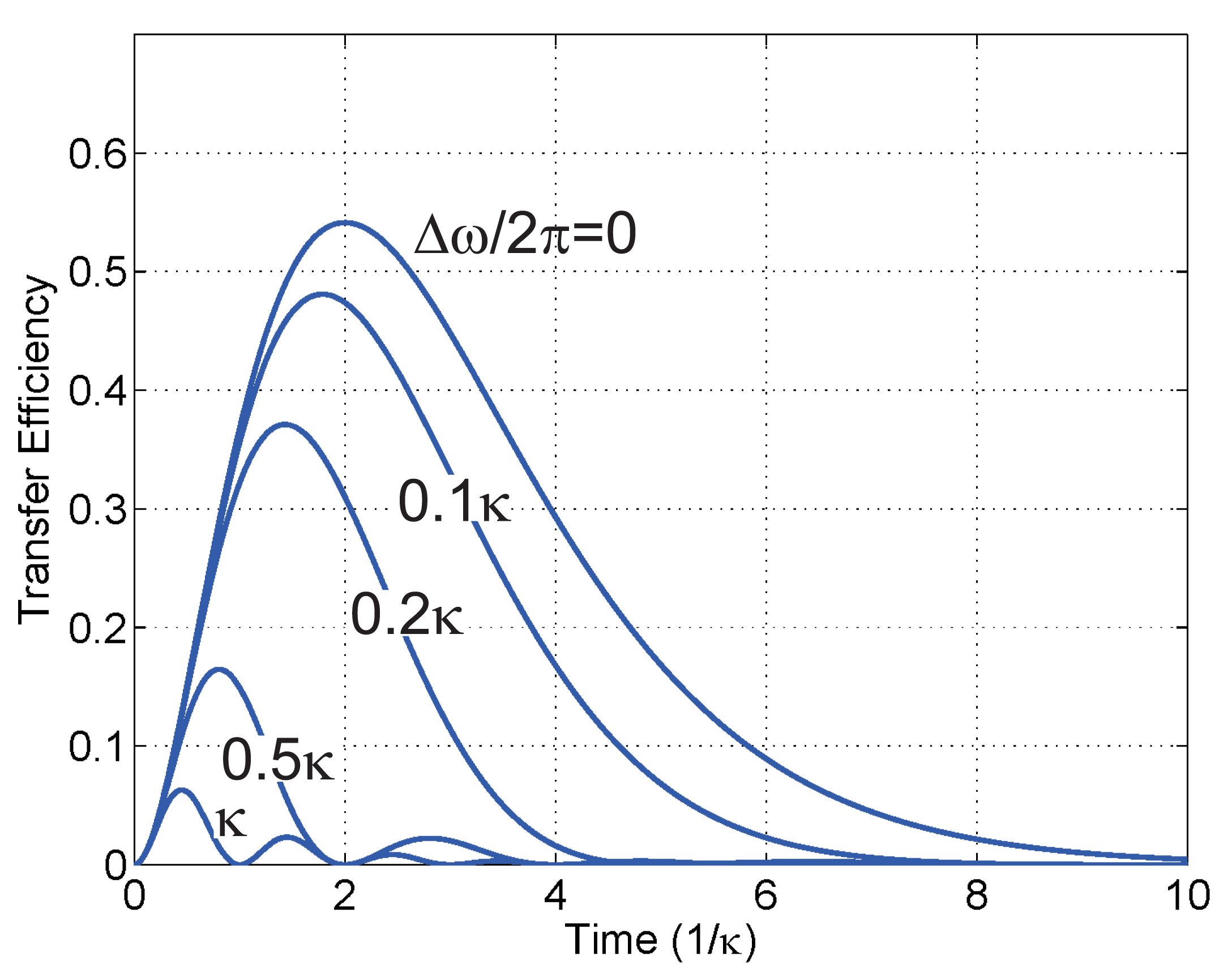}
\vspace*{-0.0in} \caption{Effect of frequency mismatch $\Delta \omega \neq 0$ on photon transfer efficiency. Here the two modes are taken to have equal decay rates $\kappa_1=\kappa_2\equiv \kappa$. The curves are labeled according to the value of $\Delta \omega/2 \pi$. The peak transfer efficiency is rather sensitive to frequency mismatch.}
\label{fig:omega}\end{figure}

In the case of temporal mode mismatch $\Delta \kappa \neq 0$, the energy transferred to mode 2 is given by
\begin{align}
E_2(t) = \frac{2V_0^2}{Z_0}\kappa_2 e^{-\kappa_2t} \frac{1}{\Delta \kappa^2} \left(e^{\Delta \kappa t/2} - 1\right)^2.
\label{eqn:tmm}
\end{align}
In Fig. \ref{fig:kappa} we plot photon transfer efficiency \textit{versus} time in the case of temporal mode mismatch. We see that peak transfer efficiency is relatively insensitive to mismatch of decay rates $\kappa$, although the time at which transfer efficiency peaks shifts away from $2/\kappa_1$, as expected.

Similarly, we can analyze the case of frequency mismatch $\Delta \omega \equiv \omega_2 - \omega_1 \neq 0$ for the case where $\Delta \kappa = 0$. We can show that
\begin{align}
E_2(t) = \frac{V_0^2}{Z_0} \kappa e^{-\kappa t} \frac{1}{\Delta \omega^2}\left[1-\cos(\Delta \omega t)\right].
\end{align}
In Fig. \ref{fig:omega}, we plot photon transfer efficiency \textit{versus} time for different amounts of frequency mismatch, assuming $\Delta \kappa = 0$. Even modest frequency mismatch degrades efficiency significantly, so that \textit{in situ} tuning of one of the cavities will be necessary for efficient transfer. Mode repulsion between the JPM and capture cavity provides the needed tunability.

\section{S6. Window Functions for Pointer State Preparation}
Window functions are typically used for pulse shaping qubit drive waveforms in order to suppress spectral content at the 1-2 transition frequency. Since JPM-based qubit state measurement relies on intensity contrast between bright and dark cavity pointer states, windowing functions were used to suppress microwave energy at the dark pointer state frequency. The Hamming window function was used for cavity pointer state preparation in our experiment. The duration of our cavity pointer state preparation pulse was 780 ns.

\section{S7. Estimates of Photon Occupation}
\begin{figure}[h!]
    \includegraphics[width=0.45\textwidth]{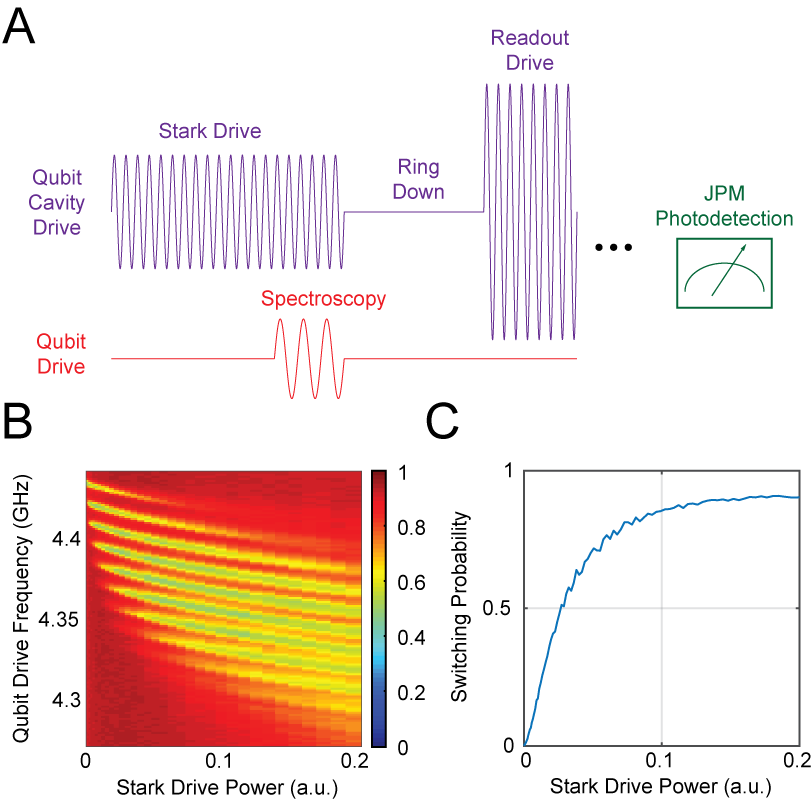}
     \caption{Stark calibration of photon occupation. \textbf{(A)} Pulse sequence used for Stark spectroscopy. Stark drive on the qubit cavity builds up a steady-state photon occupation in that mode. After steady state is reached, a spectroscopy pulse with variable frequency is applied to the qubit. Following the spectroscopic pulse, we wait for the qubit cavity to ring down. Readout drive is then applied to the qubit cavity for pointer state preparation and JPM photodetection. \textbf{(B)} Qubit spectroscopy data versus Stark drive power. Each of the sloped yellow lines corresponds to a distinct photon number state in the qubit cavity. \textbf{(C)} JPM switching probability versus Stark drive power. Here we use the Stark drive for pointer state preparation (ring down, readout drive, and spectroscopy pulse are omitted). This maps readout JPM switching probability onto photon occupation in the qubit cavity, allowing for an estimate of photon occupation of the capture cavity during bright pointer state detection. \label{fig:figS7}}
\end{figure}

JPM-based qubit state measurement relies on the transfer of cavity pointer states between the qubit and capture cavities (see Fig. \ref{fig:figS4}A). Following pointer state transfer, the state of the capture cavity is detected by the JPM. The short ($\sim10~\text{ns}$) JPM relaxation time makes it difficult to directly measure the mean photon occupation in the capture cavity \cite{Hofheinz08}; however, we can use Stark spectroscopy to calibrate photon occupation of the qubit cavity \cite{Schuster05, Schuster07}. For these experiments, we create a bright pointer state corresponding to the qubit $|0\rangle$ state in order to circumvent issues associated with qubit energy relaxation. This is in contrast with the results reported in the main text where we drove on the dressed $|1\rangle$ cavity state to create a bright pointer. The waveforms used for Stark calibration are shown in Fig. \ref{fig:figS7}A. Stark data shown in  Fig. \ref{fig:figS7}B indicate a maximum of 10 photons in the qubit cavity for the range of powers shown. Next, we map our Stark drive (leaving out the spectroscopy and readout drive pulse) onto JPM switching probability. The switching probability saturates at drive powers which correspond to a qubit cavity photon occupation of $\bar{n} \approx 8$. Using the photon transfer efficiency formula Eq. \ref{eqn:tmm} and the measured decay rates $\kappa_1~\text{and}~\kappa_2$, we can estimate a mean photon occupation in the capture cavity during JPM photodetection. In our experiment, $\kappa_2/\kappa_1 \approx 6.5$. This leads to a maximum transfer efficiency of $\approx 35$\%. From this, we estimate a capture cavity photon occupation of $\bar{n}\approx3$ during bright pointer detection.

\section{S8. Tomography Fits}
To estimate the qubit density matrix from the overdetermined tomography described in Fig. 4C of the main text, we perform a four-parameter fit to a simplistic model of the gate sequence and measurement. The model assumes perfect gates and measurement; any fidelity loss then appears as a less pure density matrix. The fit function is determined by considering an arbitrary density matrix,
\begin{equation}
\rho =
\begin{pmatrix}
1-\beta & r e^{\textit{i} \phi}\\
r e^{-\textit{i} \phi} & \beta
\end{pmatrix},
\end{equation}
which is rotated about an axis $\theta$ for a time $t$, given by
\begin{equation}
R = \exp\bigg[\frac{\textit{i} \pi}{2} \frac{t}{t_{\pi}} (\sigma_x \cos\theta + \sigma_y \sin\theta)\bigg],
\end{equation}
where $t_{\pi}$ is the $\pi$-pulse duration and $\sigma_x$ and $\sigma_y$ are the usual Pauli matrices. After the rotation, the qubit occupation is measured, $M = |1\rangle \langle 1|$.  The average occupation fit function is thus given by
\begin{equation}
P(t,\theta) = \text{Tr}(R \rho R^{\dag} M).
\end{equation}
We then fit the tomographic data to the occupation fit function, with fit parameters $\beta$, $r$, $\phi$, $t_{\pi}$, resulting in an estimate of the qubit density matrix. The extracted density matrices for the conditional tomograms shown in in Fig. 4C of the main text are
\begin{equation}
 \rho_{0} =
  \begin{pmatrix}
    0.91&0.02\\
    0.02&0.09
  \end{pmatrix}
\end{equation}
and
\begin{equation}
 \rho_{1} =
  \begin{pmatrix}
    0.31&0.01\\
    0.01&0.69
  \end{pmatrix}.
\end{equation}
Here, the subscripts 0, 1 correspond to the classical outcome of the initial qubit measurement. We then use the estimated density matrices to compute the overlap (Jozsa) fidelities $\langle\psi|\rho|\psi\rangle$, where $|\psi\rangle$ is the target state. 

\end{document}